\documentclass[12pt]{article}
 \newcommand{\be}{\begin{equation}}
 \newcommand{\ee}{\end{equation}}
 \newcommand{\bea}{\begin{eqnarray}}
 \newcommand{\eea}{\end{eqnarray}}

 \def\a{\alpha}

 \def\d{\delta}
 \def\e{\epsilon}

 \def\l{\lambda}
 
 \def\n{\nu}
 \def\o{\omega}
 \def\p{\pi}
 
 \def\r{\rho}

 \def\x{\xi}

 \def\F{\Phi}
 \def\G{\Gamma}

 \def\O{\Omega}

\begin{document}
 \renewcommand{\theequation}{\thesection.\arabic{equation}}
 \newcommand{\eqn}[1]{eq.(\ref{#1})}

 \renewcommand{\section}[1]{\addtocounter{section}{1}
 \vspace{5mm} \par \noindent
   {\bf \thesection . #1}\setcounter{subsection}{0}
   \par
    \vspace{2mm} } 
 \newcommand{\sectionsub}[1]{\addtocounter{section}{1}
 \vspace{5mm} \par \noindent
   {\bf \thesection . #1}\setcounter{subsection}{0}\par}
 \renewcommand{\subsection}[1]{\addtocounter{subsection}{1}
 \vspace{2.5mm}\par\noindent {\em \thesubsection . #1}\par
  \vspace{0.5mm} }
 \renewcommand{\thebibliography}[1]{ {\vspace{5mm}\par \noindent{\bf
 References}\par \vspace{2mm}}
 \list
  {\arabic{enumi}.}{\settowidth\labelwidth{[#1]}\leftmargin\labelwidth
  \advance\leftmargin\labelsep\addtolength{\topsep}{-4em}
  \usecounter{enumi}}
  \def\newblock{\hskip .11em plus .33em minus .07em}
  \sloppy\clubpenalty4000\widowpenalty4000
  \sfcode`\.=1000\relax \setlength{\itemsep}{-0.4em} }
 \vspace{4mm}
 \begin{center}
 {\bf Higgsed antisymmetric tensors and topological defects} \vspace{1.4cm}

 JAN TROOST  \footnote{
troost@tena4.vub.ac.be;  \,  Aspirant F.W.O.}\\
 {\em Theoretische Natuurkunde, Vrije Universiteit Brussel} \\
 {\em Pleinlaan 2, B-1050 Brussel, Belgium} \\
 \centerline{ABSTRACT}
 \end{center}
 \begin{quote}\small
We find topological defect solutions to the equations of motion of
a generalised Higgs model with antisymmetric tensor fields. These
solutions are direct higher dimensional analogues of
 the Nielsen-Olesen vortex solution 
for a gauge field in four dimensions.  
\end{quote}
 \section{Introduction}
 \noindent
It is well known that the Higgs-mechanism 
 \cite{H} 
involving a
complex scalar and a vector particle can be extended to
 antisymmetric tensor fields \cite{Q}. The formal extension
consists in a model containing a (h-1)-form, an h-form and a
scalar field with a Higgs type Lagrangian in a spacetime of
arbitrary dimension $ D=d+1 $. When the scalar field gets a vacuum
expectation value, the h-form  eats the degrees of freedom of
the (h-1)-form and  acquires a mass.
\newline
In this paper we will look for solutions to the equations of
motion of the generalised Higgs model. We will work in close
analogy to the paper of Nielsen and Olesen \cite{NO} on the
vortex-solution in four dimensions (also of use in
superconductivity). In the same approximation as in \cite{NO}, we
will find topological defect solutions that extend over $d-h-1$
dimensions.
\newline
These solutions have their importance in the study of the
different phases in antisymmetric tensor field theories and
they play a role in determining the
physical content \cite{Y} of a brane-antibrane system after
tachyon condensation \cite{S}.

 \setcounter{equation}{0}
 \section{The antisymmetric tensor Higgs model}
 We study a model in $ D=d+1 $
 dimensions with the following field content:
 an antisymmetric tensor of degree $h$, $\o_h$,  an antisymmetric tensor of
 degree $h-1$, $C_{h-1}$ and a scalar field $f$. We consider the
 following action \footnote {When using form-notation we will be
 sloppy with numerical factors, but in component form we believe to have
 every factor straight.}:
\begin{eqnarray}
{\cal S} &=& \int d^{d+1} x \, (  d \o_h  \ast d \o_h +  d f \ast
d f \nonumber
\\ & & - m (f)^2 ( d C_{h-1} + q \, \o_h) \ast (d C_{h-1}+q \,
\o_h) - U(f) ),
\end{eqnarray}
where $m(f)$ and $U(f)$ are general functions of the scalar field.
The gauge symmetries of this action are:
 \begin{eqnarray}
\d \o_h &=& d \e_{h-1} \nonumber \\ \d C_{h-1} &=& - q \, \e_{h-1}
+ d \x_{h-2}.
\end{eqnarray}
If the scalar field $f$ acquires a vacuum expectation value, it is
appropriate to use the gauge freedom  to gauge away the (h-1)-form
$C_{h-1}$ completely. A massive h-form $\o_h$ and a real scalar
will be left as physical fields, as in the ordinary
Higgs-mechanism. Later on we will make use of a specific form of
the potential:
\begin{eqnarray}
U(f)= - c_2 f^2 + c_4 f^4, \label{quartic}
\end{eqnarray}
where we took over some of the conventions of \cite{NO} for easy
comparison. Note that for $D=4$ and $h=1$ and the quartic
potential, the model matches up with the Higgs-model, where $\o_1$
is the gauge field, $C_0$ represents the phase of the complex
scalar, and $f$ its modulus. The function $m(f)$ is then given by
$m(f)^2 = f^2$.

The Lagrangian is expressed in component form as follows:
\begin{eqnarray}
\frac{{\cal L}}{\sqrt{|g|}} &=&
- \frac{1}{2 (h+1)!} ( (h+1) \partial_{[M_{h+1}} \o_{M_1 \dots
M_{h}]} )^2 \nonumber \\
& & - \frac{1}{2} (\partial_{M} f)^2 - m(f)^2 \frac{1}{2 h!}
( h \, \partial_{[M_{h}} C_{M_1 \dots
M_{h-1}]} + q \, \o_{M_1 \dots M_h} )^2     - U(f)
\end{eqnarray}
The equations of motion corresponding to this Lagrangian are:
\begin{eqnarray}
0 &=& \frac{1}{\sqrt{|g|}} \partial_{M_{h+1}} \sqrt{|g|} (d
\o)^{M_{h+1} M_1 \dots M_h} \nonumber \\ & & - m(f)^2 q (  h \,
\partial^{[M_h} C^{M_1 \dots M_{h-1}]} + q \, \o^{M_1 \dots M_h} )  \\ 0 &=&
\frac{1}{\sqrt{|g|}} \partial_M (\sqrt{|g|} \partial^M f)
\nonumber \\ & & - \frac{1}{h !} m(f) m(f)' (h \, \partial_{[M_h}
C_{M_1 \dots M_{h-1} ]} + q \, \o_{M_1 \dots M_h} )^2 - U(f) '  \\
0 &=& \partial_{M_1} ( \sqrt{|g|} m(f)^2 (h \, \partial^{[M_h}
C^{M_1 \dots M_{h-1}]} + q \, \o^{M_1 \dots M_h}) )  ,
\end{eqnarray}
where a prime denotes differentiation with respect to $f$. The
last equation is the equation of motion corresponding to the
$C_{h-1}$ form. It merely states that the h-form $\o_h$ couples to
a conserved current
\begin{eqnarray}
j^{M_1 \dots M_h} &=& q \sqrt{|g|} m(f)^2 (h \, \partial^{[M_h}
C^{M_1 \dots M_{h-1}]} + q \, \o^{M_1 \dots M_h}).
\end{eqnarray}

 \setcounter{equation}{0}
\section{Ansatz and solution}

In close analogy to \cite{NO}, we will  look for a topological
defect solution of dimension $d-h-1$. The (h+1)-form field
strength will measure the number of topological defects passing
through a $h+1$ dimensional plane perpendicular to the 
defects. We can define a magnetic flux $\F$ flowing through a
(h+1)-dimensional ball and calculate it in terms of the (h-1)-form
field strength
\begin{eqnarray}
\F & \equiv & \int_{B^{h+1}} d \o_{h} \nonumber \\
 & =& \int_{S^h} \o_h \nonumber \\
 &=& - \frac{1}{q} \int_{S^h}  d C_{h-1}, \label{flux}
\end{eqnarray}
where we have used the fact that there is no current over the
h-sphere that is the boundary of the (h+1)-ball. \footnote{
For the case $h=1$ it is clear that the magnetic flux is
quantized \cite{NO}.
For $h = 2$ see \cite{Y}. To us it seems that in the case $ h \ge
2 $ you could suppose the existence of an electric charge for the
(h-1)-form to have quantization of the magnetic flux for the
h-form, reasoning along the lines of \cite{N}.}

We  consider an ansatz with $SO(h+1) \times \mbox{Poincar\'e} \,
(d-h-1,1)$ symmetry. We use the following coordinates adapted to
the symmetry: $(r, \phi, \theta_1, \dots, \theta_{h-1}, t,z_1,
\dots, z_{d-h-1})$ The ansatz reads in these coordinates:
\begin{eqnarray}
\o_h &=& |\o(r)| r^h d \O_h \\ C^{\pm}_{h-1} &=&  k_{h-1} ( \pm
l_{h-1} + f_{h-1} (\theta_{h-1})) \, d \O_{h-1}
\end{eqnarray}
where $ d \O_h $ denotes the volumeform of the h-sphere
 with volume $s_h$. Moreover we take the constants $k_{h-1}$ and $l_{h-1}$
to be
\begin{eqnarray}
k_{h-1} &=& \frac{2 \p }{s_h} \nonumber \\
        &=& \p^{\frac{-(h-1)}{2}} \G(\frac{h+1}{2}) \label{k} \\
l_{h-1} &=& \frac{\sqrt{\p}}{2}
\frac{\G(\frac{h}{2})}{\G(\frac{h+1}{2})},
\end{eqnarray}
and the function $f_{h-1}$ satisfies
 \begin{eqnarray}
\frac{d}{d \theta_{h-1}} f (\theta_{h-1}) &=& \sin^{h-1}
\theta_{h-1},
\end{eqnarray} following \cite{N}. From these formulae we easily derive:
\begin{eqnarray}
d \o_h &=& \partial_r(|\o| r^h) \, dr \wedge d\O_h \\
d C_{h-1} &=&  k_{h-1} \sin^{h-1} \theta_{h-1} d \theta_{h-1} \wedge
d \O_{h-1} \nonumber \\
& = &  k_{h-1} \, d \O_h
\end{eqnarray}
The (h-1)-form ansatz is chosen such
that from the formula for the flux (\ref{flux}), we can conclude
that there will be a topological defect in the $(t,z_1, \dots
z_{d-h-1})$
 direction.
Filling in the ansatz in the equations of motion results in the following
set of differential equations:
\begin{eqnarray}
0 &=& \frac{1}{r^h} \partial_r (r^h \partial_r f) \nonumber \\ & &
- m(f) m(f)' ( \frac{ k_{h-1}}{r^h} + q |\o|)^2 - U(f)'
\label{feom}\\ 0 & = & \partial_r (\frac{1}{r^h} \partial_r (| \o
| r^h) ) - m(f)^2 q \,(q |\o| + \frac{ k_{h-1}}{r^h})
\label{oeom}
\end{eqnarray}
Following \cite{NO}, we consider the situation in which $f$ tends
to a constant value at  infinity (transverse to the topological
defect). In that approximation, we can solve (\ref{oeom}) for
$|\o|$ in terms of modified Bessel functions \footnote{Details of
the standard manipulations are in the first appendix.}:
\begin{eqnarray}
| \o | &=& -\frac{ k_{h-1}}{q} \frac{1}{r^h} + \frac{c}{q} (q m
r)^{-\n+1} K_{\n} ( q m r )   \label{sol}
\end{eqnarray}
where the index of the modified Bessel function is given by $ \n =
\frac{h+1}{2} $ and $c$ is an integration constant. The magnetic
field strength becomes:
\begin{eqnarray}
|H| & \equiv & \frac{1}{r^h} \partial_r (r ^h |\o|) \nonumber \\
 &=& c m (q m r)^{-\n+1} K_{\n-1} (q m r)
\end{eqnarray}
where we have used a property of the modified Bessel function
given in the second appendix (\ref{der}). The asymptotic behavior
of the magnetic field is then (\ref{as}):
\begin{eqnarray}
 |H|_{r  \to  \infty} &=& c \, \sqrt{\frac{\p m}{2 q r}} (q m
r)^{-\n+1} e^{-q m r} + \dots
\end{eqnarray}
We can  define a characteristic length measuring the distance
over which the magnetic field differs appreciably from zero:
\begin{eqnarray}
\l & \equiv & \frac{1}{q m}
\end{eqnarray}
This is the analog of the penetration depth in superconductivity.

We turn now to the other equation of motion (\ref{feom}), the one
corresponding to the scalar field $f$ . We will restrict from this
point on to the special case of a quartic Higgs potential
(\ref{quartic}). If we assume that the deviation of the h-form
from $ -\frac{ k_{h-1} }{q r^h}$ is negligable compared to the
effect of a steep potential, in other words, if the coefficients
$c_2$ and $c_4$ are large, then (\ref{feom}) is approximately
satisfied if $f$ takes the constant value that minimizes the
potential:
\begin{eqnarray}
<f> & \equiv & v \nonumber \\
&=& \sqrt{\frac{ c_2}{2 c_4}}
\end{eqnarray}
To get an idea of the position dependence of $f$ we consider
fluctuations around its vacuum expectation value \cite{NO}:
\begin{eqnarray}
f &=& v + \r (r)
\end{eqnarray}
At infinity we find an approximate solution
\begin{eqnarray}
\r(r) & \approx & e^{-\sqrt{4 c_2} r}
\end{eqnarray}
giving rise to a new characteristic length,
\begin{eqnarray}
\x &=& \frac{1}{\sqrt{4 c_2}},
\end{eqnarray}
the generalisation of the correlation length in superconductivity.
It is nothing but the inverse of the mass of the Higgs-particle,
and it measures the distance over which $f$ differs appreciably
form its vacuum expectation value. All of this to spell out that
the behavior of the fields is analogous to the well known special
case of the Nielsen-Olesen vortex-solution, due to the general
properties of the modified Bessel functions. We further remark
that to have a clear corelike topological defect of dimension
$d-h-1$, we need $\x$ and $\l$ both small.  For $ r  << \l $ we
require that the magnetic flux $\F = V(S^h) |\o| $ vanish, fixing
the integration constant to:
\begin{eqnarray}
c  &=&   (2 \p)^{-\n+1} (q m)^{h} ,
\end{eqnarray}
where we made use of formula (\ref{zero}) in the appendix,
(\ref{k}) and (\ref{sol}). This finishes the discussion of the
approximate solution.

 \setcounter{equation}{0}
\section{A remark}

For some applications it is useful to have an estimate of the
energy density of the $d-h-1$-dimensional topological defect. In
\cite{NO} the idea was to match up the energy density of the
vortex with the string tension of the dual string model. The
reasoning was that the string model would be a good effective
description of the field theory in a regime where the vortex
solution becomes the most important classical solution to the
action \cite{NO}. Then you can link the field theory parameters to
the string tension, $\frac{1}{2 \p \a'}$. Following \cite{Y} we
can study the special case of the membrane-like solution in
$D=5+1$ and with $h=2$ and try to match its energy density to the
M2-brane tension. We only make a rough analysis, indicating that
this is less straightforward.

Treating the mass $m$ as a constant, the magnetic contribution to
the energy density is in the general case \cite{NO}
:
\begin{eqnarray}
E_m& \approx &\frac{s_h}{2} \int_0^{\infty} |H|^2  r^h dr
\nonumber
\\  & \approx &
 \frac{   s_h }{2 (2 \p)^{2 \n-2}} q^{h-1} m^{h+1}
 \int_0^{\infty}    K_{\n-1} (z) ^2 z \, dz
\end{eqnarray}
From the asymptotic behavior of the modified Bessel functions
(\ref{as}), it is clear that the integrand converges fast enough
at infinity, but near $z=0$ (\ref{zero}), the integrand behaves as
$z^{-h+2}$. For a finite result we thus need $h < 3$. Then the
integral is of order unity and the magnetic energy density of
order
\begin{eqnarray}
E_m & \approx & q^{h-1} m^{h+1}
\end{eqnarray}

The contribution to the energy density of the topological defect
due to the scalar field $f$ can roughly be approximated by
\cite{NO}:
\begin{eqnarray}
E_f  & \approx & \x^{h+1} c_2 v^2 \nonumber \\ & \approx &
\frac{c_2^{\frac{3-h}{2}} }{c_4}
\end{eqnarray}
We note that for $h=1$ and $m=v=\sqrt{\frac{c_2}{2 c_4}}$ the two
estimates match \cite{NO}. For other values of $h$ like $h=2$ the
naive analysis gives an ambiguous result. We will not pursue this
here. Notice though, that our sketchy analysis of the asymptotic
behavior of the magnetic field and the energy density is finer
than the one in \cite{Y}.

 \section{Conclusions}

We studied  Higgsed antisymmetric tensor field theories.
Specifically we searched for and found topological defect
solutions, generalizing the work by Nielsen and Olesen on vortex
solutions in four dimensions \cite{NO}. We wrote down the
approximate behavior of the fields at infinity in terms of
modified Bessel functions, making use of the assumption that the
scalar field there reaches a constant vacuum expectation value. We
briefly indicated the possibility of identifying the energy
density of the topological defect solutions in terms  of field
theory parameters, with possible other descriptions of the same
objects. We hope that these explicit solutions may be of benefit
to a study of the phases of antisymmetric tensor theories and to
the study of brane anti-brane systems in string theory.

 \vspace{1cm}

 \noindent {\bf Acknowledgments}:
Thanks are due to Fernando Quevedo for advice and to Alex Sevrin
and Walter Troost for useful discussions.
 \newpage
 \appendix
 \noindent
  {\bf APPENDIX}
   \setcounter{equation}{0}
 \section{Solution of the differential equation}
\noindent We solve the differential equation (\ref{oeom}):
\begin{eqnarray}
0 & = & \partial_r (\frac{1}{r^h} \partial_r (| \o | r^h) ) - m(f)^2 q \,
(q \, |\o| + \frac{ k_{h-1}}{r^h}) \nonumber
\end{eqnarray}
for constant $f$ by standard techniques.
First, we define a new unknown function, $\O$, and a new variable $z$ :
\begin{eqnarray}
\O & \equiv & |\o| + \frac{ k_{h-1}}{q} \frac{1}{r^h} \\
z & \equiv & q m r
\end{eqnarray}
In terms of the new variables the differential equation takes the form:
\begin{eqnarray}
 \partial_z \partial_z \O + \frac{h}{z} \partial_z \O
&=& \O + \frac{h}{z^2} \O
\end{eqnarray}
To bring this to a well known form, we define still another
function $ X = \O z^{\frac{h-1}{2}} $, in terms of which the
differential equation reads:
\begin{eqnarray}
 \partial_z \partial_z X + \frac{1}{z} \partial_z X
&=& (1+ \frac{ \n^2}{z^2} ) X
\end{eqnarray}
where $ \n= \frac{h+1}{2} $.
Excluding the solutions which blow up at infinity, we find the following
standard solution for $X$ in terms of modified Bessel functions,
including an integration constant:
\begin{eqnarray}
X &=& \frac{c}{q} K_{\n} (z)
\end{eqnarray}
Returning to the original variables gives:
\begin{eqnarray}
\O &=& (qmr)^{-\n+1} \frac{c}{q} K_{\n} (q m r) \\ |\o| &=&
-\frac{ k_{h-1}}{q} \frac{1}{r^h} +  \frac{c}{q} (qmr)^{-\n +1}
K_{\n} (q m r)
\end{eqnarray}

 \setcounter{equation}{0}

\section{Some properties of modified Bessel functions}
For easy reference we list here the asymptotic behavior of the
modified Bessel functions $K_{\n}$ and a property of the
derivative that we will need in the body of the text:
\begin{eqnarray}
z  \to  0 & & \nonumber \\ K_0 (z) & \approx & -\log z \nonumber
\\ K_{\n}(z) & \approx & \frac{1}{2} \G (\n) (\frac{1}{2} z)^{-\n}
\label{zero} \\ & & \nonumber \\ z \to \infty & & \nonumber \\
K_{\n} (z) & \approx & \sqrt{\frac{\p}{2z}} e^{-z}  \label{as} \\
& & \nonumber
\\ K_{0} (z)' &=& - K_1 (z) \nonumber \\
               \label{der} \frac{1}{z}
\frac{d}{dz} (z^{\n} K_{\n} (z))  &=& z^{\n-1} K_{\n-1} (z)
\end{eqnarray}

 \end{document}